\documentclass[aps,prl,twocolumn,superscriptaddress,10pt]{revtex4-2}
\usepackage[english]{babel}
\usepackage[utf8]{inputenc}
\usepackage[T1]{fontenc}
\usepackage{amsmath,amssymb}
\usepackage{physics}
\usepackage{graphicx}
\usepackage[pdftitle={Machine Learning the GW Self Energy Operator for Real Materials}, pdfauthor={Xian Xu, Bowen Hou, Chengyan Zhang, Diana Y. Qiu}]{hyperref}

\graphicspath{{figures/}}

\begin{document}

\title{Machine Learning the $GW$ Self-Energy Operator for Real Materials}

\author{Xian Xu}
\affiliation{Department of Applied Physics, Yale University, New Haven, CT, 06511, USA}
\author{Bowen Hou}
\affiliation{Department of Materials Science, Yale University, New Haven, CT, 06511, USA}
\author{Chengyan Zhang}
\affiliation{Department of Applied Physics, Yale University, New Haven, CT, 06511, USA}
\author{Diana Y. Qiu}
\email{diana.qiu@yale.edu}
\affiliation{Department of Materials Science, Yale University, New Haven, CT, 06511, USA}

\date{\today}

\begin{abstract}
Machine learning of $GW$ in solids has largely focused on prediction of scalar quasiparticle energy corrections. Here, we introduce a phase-aware many-body transformer that predicts the full $\mathbf{k}$-resolved complex self-energy matrix. As a proof of principle, we apply the model to predict quasiparticle energies before and after diagonalization of the quasiparticle Hamiltonian, obtaining a test set mean absolute error (MAE) of 29 meV for a dataset of 1745 2D materials in the static COHSEX approximation and 60 meV for a more limited dataset of 433 materials calculated within a plasmon-pole model. COHSEX pretraining followed by finetuning reduces this MAE to 42 meV. The predicted matrices also accurately reproduce QP wavefunctions.
\end{abstract}

\maketitle

Machine learning (ML) has become a powerful tool for accelerating first-principles calculations of materials' ground-state properties~\cite{Carleo2019RMP}, ranging from neural-network interatomic potentials that can reach the accuracy of density functional theory (DFT) at a small fraction of its computational cost~\cite{Behler2007, Batzner2022NequIP} to graph neural networks that can predict materials properties and independent-particle Hamiltonians directly from the atomic structure~\cite{Xie2018CGCNN,Li2022DeepH,Gong2023DeepHE3}. Most of these approaches are built on DFT, the workhorse of ground-state electronic structure and the basis of large-scale materials databases~\cite{Jain2013MaterialsProject}. However, Kohn-Sham (KS) eigenvalues~\cite{Hohenberg1964,Kohn1965} belong, in principle, to an auxiliary system of noninteracting particles and do not describe the quasiparticle (QP) energies measured in photoemission and tunneling spectroscopy. Within first-principles many-body perturbation theory, the $GW$ approximation is the standard approach for predicting QP properties of solids~\cite{Hedin1965,Hybertsen1986}. Within the $GW$ approximation, the electron correlation is encoded in the nonlocal self energy operator $\Sigma=iGW$, where $G$ is the single-particle Green's function and $W$ is the screened Coulomb interaction. $GW$ calculations are, however, typically orders of magnitude more expensive than DFT~\cite{Deslippe2012BerkeleyGW}.

The high computational cost of $GW$ has motivated the development of ML surrogates for QP energies and self-energy corrections. Early machine-learning approaches to $GW$ focused on accelerating intermediate quantities such as the dielectric screening~\cite{Dong2021Screening,Zauchner2023}. Most subsequent work has directly targeted QP energy corrections, corresponding to the diagonal self-energy in the Kohn-Sham (KS) basis. These approaches have been used to transfer $GW$ accuracy across configurations and system sizes for individual materials~\cite{Pant2026,Abdelghany2025HTGW,Gong2026DeepHGW}, as well as across chemical space for molecules~\cite{Caylak2021,Westermayr2021} and solids~\cite{Knosgaard2022,Hou2024,Hou2025MBFormer}.
More recently, machine learning has been extended to the off-diagonal self-energy matrix: Venturella and co-workers demonstrated transferable prediction across chemically diverse molecules~\cite{Venturella2024,Venturella2025}, while applications to solids have so far been limited to model Hamiltonians~\cite{Zhu2025SigmaAttention} and $\Gamma$-point only models for simple solids like diamond and silicon~\cite{Dong2024Green}. 
A $k$-resolved prediction of the complex $GW$ self-energy matrix across chemically diverse solids has not yet been demonstrated.

Going beyond a diagonal one-shot $G_0W_0$ correction, the self-energy matrix elements $\Sigma_{mn\mathbf{k}}=\mel{m\mathbf{k}}{\Sigma}{n\mathbf{k}}$ are needed to obtain QP wavefunctions~\cite{Tang2022Water}, enable quasiparticle self-consistent $GW$ (QSGW) calculations~\cite{Faleev2004,vanSchilfgaarde2006QSGW}, and provide the QP basis required for downstream treatments of electron-phonon interactions~\cite{Li2019GWPT} and nonequilibrium dynamics~\cite{Chan2021TDaGW,ChangLee2024,Hou2025LowRank}.
At the mean-field level, the analogous step from learning eigenvalues to learning the full Hamiltonian is already well established~\cite{Schutt2019SchNOrb, Nigam2022Hamiltonian, Li2022DeepH, Zhang2022ACEH, Gong2023DeepHE3, Zhong2023HamGNN, Tang2024DeepHHybrid}. 
However, directly learning $\Sigma_{mn\mathbf{k}}$ for solids in the Bloch basis poses two additional challenges. First, $\Sigma_{mn\mathbf{k}}$ is complex-valued and gauge-covariant. Under the arbitrary $U(1)$ phase freedom of each Bloch state, $\psi_{n\mathbf{k}}\to e^{i\theta_{n\mathbf{k}}}\psi_{n\mathbf{k}}$, the matrix elements corotate as
\begin{equation}
\Sigma_{mn\mathbf{k}} \to e^{\,i(\theta_{n\mathbf{k}}-\theta_{m\mathbf{k}})}\,\Sigma_{mn\mathbf{k}},
\label{eq:gauge}
\end{equation}
so a faithful surrogate must retain the relative interband phase.
Second, small errors in individual matrix elements can be amplified by diagonalization. Accurate Hamiltonian learning therefore demands substantially greater element-wise fidelity than regression of scalar QP energies.

In this Letter, we overcome these challenges and demonstrate the $\mathbf{k}$-resolved prediction of the complex $GW$ self energy matrix across chemically distinct 2D materials. In our previous work, we introduced MBFormer~\cite{Hou2025MBFormer}, a transformer-based framework for predicting many-body corrections from mean-field electronic states in which attention~\cite{Vaswani2017Attention} uses relationships among mean-field states to predict their many-body renormalization. Here, we extend MBFormer from diagonal QP corrections to the full self energy matrix. The key advance is a phase-aware autoencoder that compresses both the amplitudes and phases of the complex KS wavefunctions, retaining the relative interband phase required to predict off-diagonal matrix elements. Additionally, we improve the accuracy, transferability, and efficiency of the model by introducing 1) joint training of the autoencoder and transformer, 2) a common physical-momentum representation of the wavefunctions, 3) gauge augmentation, and 4) factorized attention over momentum and band-pair degrees of freedom. 
We demonstrate this approach for both the static Coulomb-hole screened exchange (COHSEX)~\cite{Hedin1965,Hybertsen1986} and the dynamically screened generalized plasmon-pole (GPP)~\cite{Hybertsen1986,Deslippe2012BerkeleyGW} self energies. The resulting models achieve state-of-the-art QP-energy mean absolute errors (MAE) of 29 meV for COHSEX and 60 meV for GPP, retain their accuracy upon diagonalization, and reproduce  QP wavefunctions. Joint training further reduces the QP error and makes many-body information more accessible in the wavefunction latent space.

We start by building two $GW$ datasets with the BerkeleyGW code~\cite{Deslippe2012BerkeleyGW} on top of plane-wave DFT from Quantum ESPRESSO~\cite{Giannozzi2009}. First, we calculate static COHSEX self-energy matrices for 1745 2D semiconductors whose structures are taken from the  C2DB database~\cite{Haastrup2018C2DB, Gjerding2021C2DB}. Because COHSEX retains the matrix structure of the $GW$ self energy while being substantially cheaper to compute, it provides a practical setting for developing the model and testing its transferability across materials. Then, we apply the same framework to GPP self energy matrices for 433 hexagonal materials to assess its performance with dynamical screening.  For each dataset, 90\% of the materials are used for training and 10\% are withheld for testing. Details of dataset generation are given in the Supplemental Information (SI).

\begin{figure*}
\centering
\includegraphics[width=0.95\textwidth]{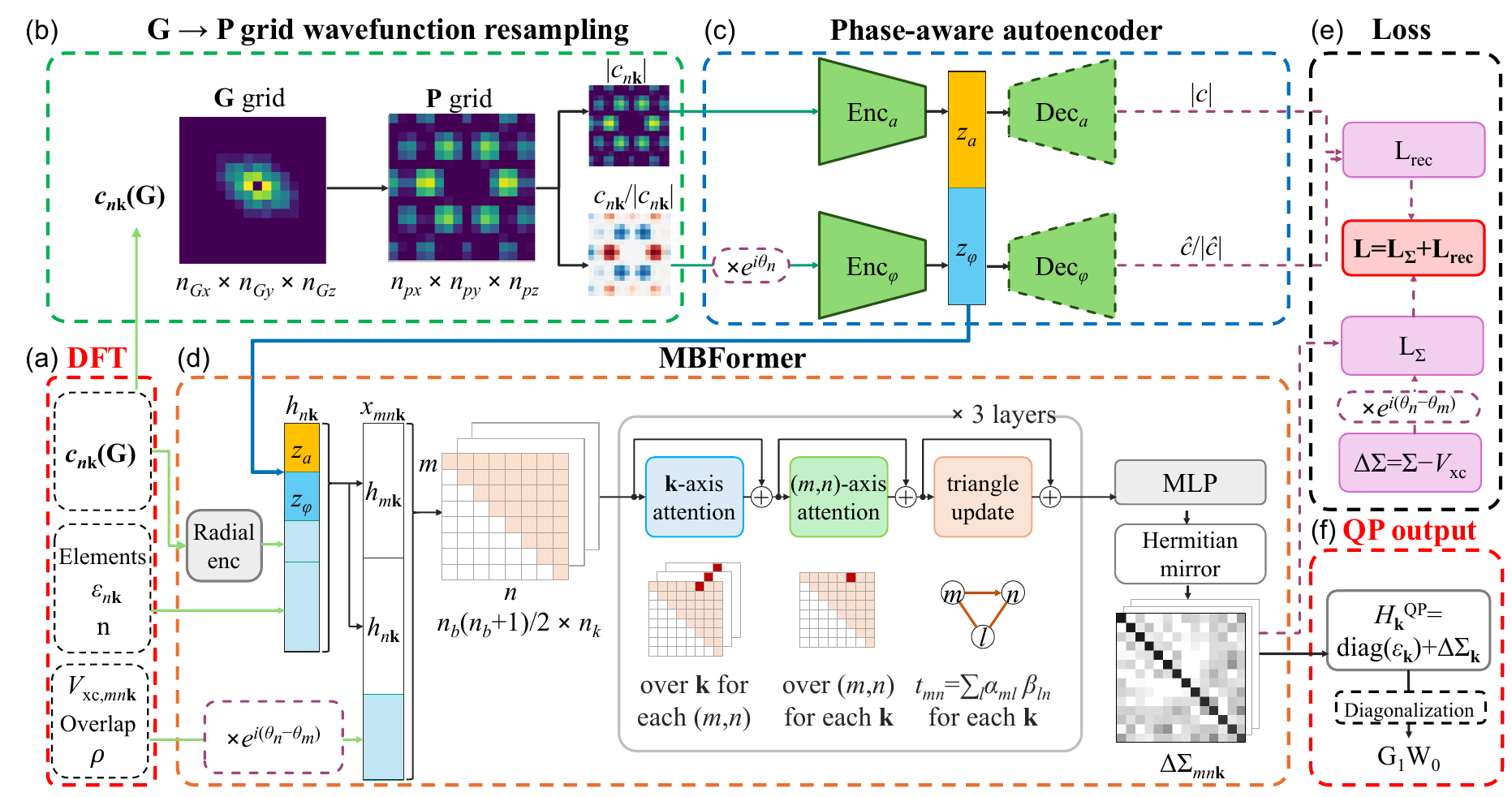}
\caption{\textbf{Architecture.}
\textbf{a}, Inputs from DFT.
\textbf{b}, The plane-wave coefficients $c_{n\mathbf{k}}(\mathbf{G})$ of each KS state are resampled from the G grid ($n_{Gx}\times n_{Gy}\times n_{Gz}$) to a grid of physical momenta $\mathbf{k}+\mathbf{G}$ (i.e. the P grid of dimension $n_{px}\times n_{py}\times n_{pz}$) and split into amplitude and phase.
\textbf{c}, The phase-aware autoencoder compresses them separately into the latents $z_a$ and $z_\phi$ and reconstructs the amplitude and the phase.
\textbf{d}, In MBFormer, the state vector $h_{n\mathbf{k}}$ combines $z_a$, $z_\phi$, a radial encoding of $\lvert c_{n\mathbf{k}}\rvert^{2}$, the element decomposition, $\varepsilon_{n\mathbf{k}}$, and the band indices. The pair tokens $x_{mn\mathbf{k}}$ ($m\le n$) combine $h_{m\mathbf{k}}$, $h_{n\mathbf{k}}$, $V_{xc,mn\mathbf{k}}$, overlaps, and density. Three layers of $\mathbf{k}$-axis attention, $(m,n)$-axis attention, and a triangle update, followed by an MLP head and a Hermitian mirror, give $\Delta\Sigma_{mn\mathbf{k}}=\Sigma_{mn\mathbf{k}}-V_{xc,mn\mathbf{k}}$.
\textbf{e}, $\mathcal{L}_{\mathrm{rec}}$ and $\mathcal{L}_{\Sigma}$ are optimized jointly. $e^{i\theta_n}$ and $e^{i(\theta_n-\theta_m)}$ denote the random gauge transformation of Eq.~\eqref{eq:gauge} applied during training.
\textbf{f}, Diagonalizing $H^{\mathrm{QP}}_{\mathbf{k}}=\mathrm{diag}(\varepsilon_{\mathbf{k}})+\Delta\Sigma_{\mathbf{k}}$ yields the $G_1W_0$ QP energies and wavefunctions.
}
\label{fig1}
\end{figure*}

Fig.~\ref{fig1} summarizes the model architecture. 
We start with an explicit KS DFT calculation. Then, we construct a common, phase-aware representation of each KS state. For each KS Bloch state, we start by mapping the planewave coefficients $c_{n\mathbf{k}}(\mathbf{G})$ to a physical momentum $\mathbf{k}+\mathbf{G}$ and resample it with a fixed Gaussian kernel onto a common $16\times16\times36$ grid of momenta (P grid in Fig.~\ref{fig1}b).
This gives the coefficient array a consistent physical meaning across materials, so that its coordinates correspond to the same physical momenta rather than material-dependent reciprocal-lattice indices. Having established this common momentum-space representation, we next decompose the coefficients into amplitude and phase components, allowing us to go beyond the amplitude-only representation used in our previous work~\cite{Hou2024,Hou2025MBFormer}.
We write the coefficients as $c=\lvert c\rvert e^{i\phi}$ and encode and decode the amplitude and phase on separate branches (Fig.~\ref{fig1}c). The phase branch predicts $(\cos\hat{\phi},\sin\hat{\phi})$ rather than $\hat{\phi}$ directly, avoiding the artificial discontinuity between phases near 0 and $2\pi$. Both branches are convolutional networks with ConvNeXt-type residual blocks over the inplane momenta. Each state is then compressed into separate amplitude and phase latents, $z_a$ and $z_\phi$, with a combined dimensionality of 280, corresponding to a compression of nearly two orders of magnitude.

The compressed states are then fed into a factorized many-body transformer (Fig.~\ref{fig1}d) whose tokens are band pairs $(m,n)$ with $m\le n$, since the remaining elements follow from Hermiticity.
We construct the inputs to mirror the inputs of an explicit $GW$ calculation.
For every $\mathbf{k}$ point, the token $x_{mn\mathbf{k}}$ concatenates the state vectors $h_{m\mathbf{k}}$ and $h_{n\mathbf{k}}$ with the exchange-correlation matrix element $\mel{m\mathbf{k}}{V_{xc}}{n\mathbf{k}}$, pair-wise overlap descriptors (see SI), and charge densities (Figs.~\ref{fig1}a,d). Each state vector combines the latents $z_a$ and $z_\phi$ with additional physical descriptors of the state (see SI).
Because the additional pair-wise descriptors in $x_{mn\mathbf{k}}$ connect the same two KS states as $\Sigma_{mn\mathbf{k}}$, they carry the same relative interband phase and provide a gauge-covariant reference for the complex output.

During training, we apply exact $U(1)$ gauge augmentation. Random phases $e^{i\theta_{n\mathbf{k}}}$ are applied to the KS wavefunctions (Fig.~\ref{fig1}c), and the label $\Sigma$ and the pair features  (Figs.~\ref{fig1}d,e) are corotated according to Eq.~\eqref{eq:gauge}. While the autoencoder has a learned gauge-covariance, the transformer is not explicitly gauge-equivariant. Instead, the augmentation teaches the network the required covariance~\cite{Zhang2026}. 

The large number of band-pair tokens makes full attention over all $m,n,\mathbf{k}$ combinations computationally expensive. We reduce this cost and guide the band coupling by taking advantage of the physical structure of the $GW$ self energy.
In the KS basis, the $GW$ self energy has the form
\begin{equation}
\Sigma_{mn\mathbf{k}}(\omega)
\sim
\sum_{\mathbf{q},l}\int d\omega'\,
\frac{
M^\dagger_{lm}(\mathbf{k},\mathbf{q})\,
W(\mathbf{q},\omega')\,
M_{ln}(\mathbf{k},\mathbf{q})
}{
\omega-\omega'-\epsilon_{l,\mathbf{k}-\mathbf{q}}+i\eta
},
\label{eq:gw_sigma}
\end{equation}
where $l$ runs over intermediate bands, $\mathbf{q}$ over momentum transfers, $\mathbf{M}_{ln}$ is the vector over $\mathbf{G}$ of plane-wave matrix elements $\mel{l\mathbf{k}-\mathbf{q}}{e^{-i(\mathbf{q}+\mathbf{G})\cdot\mathbf{r}}}{n\mathbf{k}}$, and $\mathbf{W}$ is the screened Coulomb interaction.
Because the sums over bands and over momenta are separate, we factorize attention along the $\mathbf{k}$-point and band-pair axes (Fig.~\ref{fig1}d).
For the present system size of 25 $\mathbf{k}$ points and 105 band-pair tokens per $\mathbf{k}$ point, this reduces the attention cost by a factor of roughly $20$.
Eq.~\eqref{eq:gw_sigma} also couples $m$ and $n$ through the product $M^{*}_{lm}M_{ln}$. Each transformer layer therefore ends with a gated triangle update~\cite{Jumper2021AlphaFold}, which adds to each token $x_{mn}$ a learned map of $t_{mn}=\sum_{l}\alpha_{ml}\beta_{ln}$ (Fig.~\ref{fig1}d), where $\alpha$ and $\beta$ are learned projections of the pair tokens at the same $\mathbf{k}$, and $l$ runs over the bands included in the model so that each pair $(m,n)$ receives information from all pairs $(m,l)$ and $(l,n)$, in analogy to the contraction over intermediate bands in Eq.~\eqref{eq:gw_sigma}.
A lightweight head then maps each token to $(\mathrm{Re}\,\Delta\Sigma_{mn\mathbf{k}},\mathrm{Im}\,\Delta\Sigma_{mn\mathbf{k}})$ with $\Delta\Sigma=\Sigma-V_{xc}$. Diagonalizing the QP Hamiltonian $H^{\mathrm{QP}}_{\mathbf{k}}=\mathrm{diag}(\varepsilon_{\mathbf{k}})+\Delta\Sigma_{\mathbf{k}}$ then yields updated QP energies and wavefunctions (Fig.~\ref{fig1}f).

Finally, the autoencoder and the transformer are trained jointly using the composite loss (Fig.~\ref{fig1}e)
\begin{equation}
\mathcal{L}=\mathcal{L}_{\Sigma}+\mathcal{L}_{\mathrm{rec}},
\label{eq:loss}
\end{equation}
where $\mathcal{L}_{\Sigma}$ is the error of the complex matrix elements of $\Delta\Sigma$ and $\mathcal{L}_{\mathrm{rec}}$ is the error of the reconstructed amplitude and phase on the P grid. Because the transformer operates directly on the autoencoder latent, the self energy loss also updates the encoder, allowing the representation to adapt to the many-body prediction task rather than reconstruction alone.
\begin{figure}
\centering
\includegraphics[width=\columnwidth]{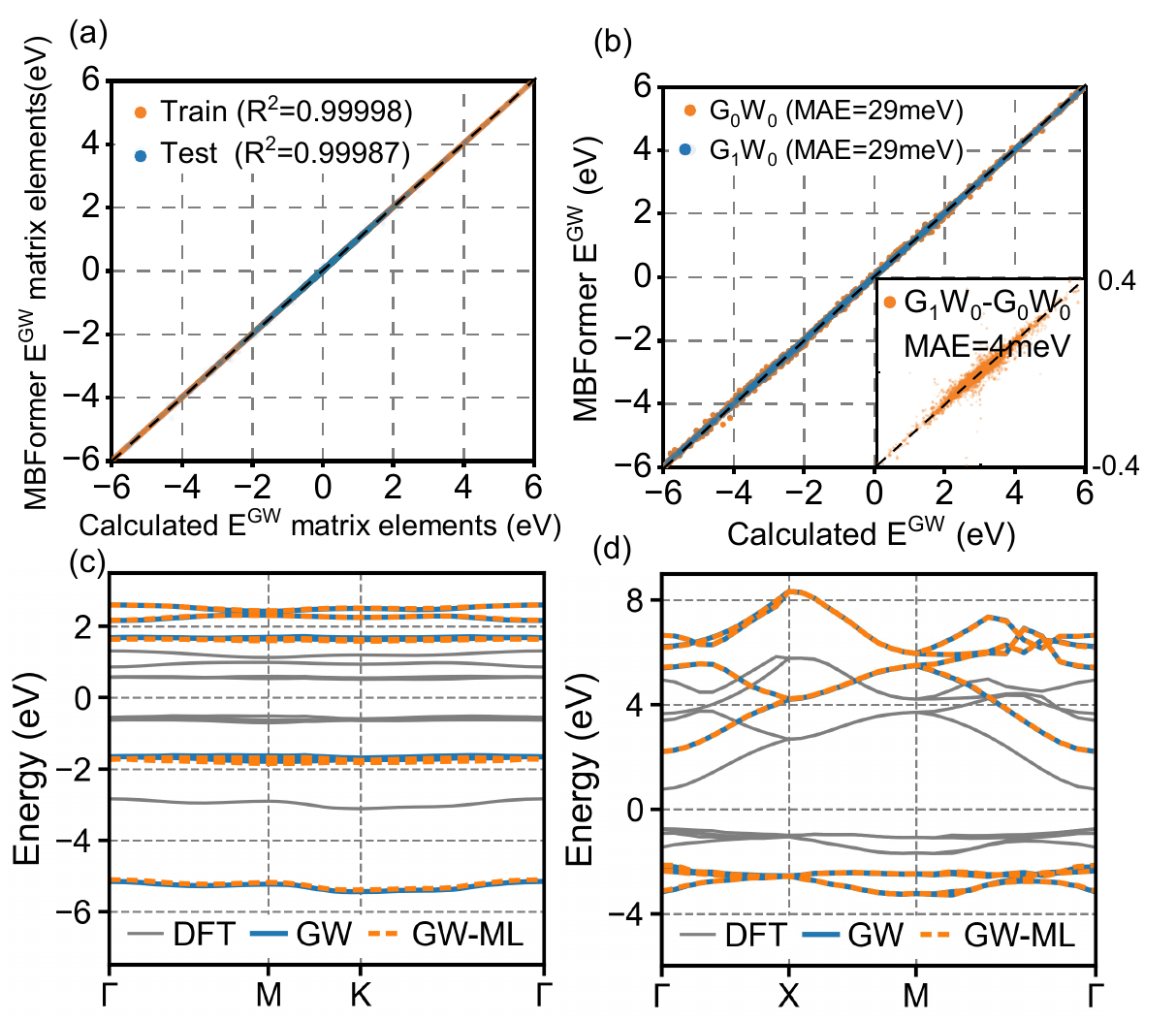}
\caption{\textbf{Prediction of static COHSEX self energy matrices.}
\textbf{a}, Predicted vs.\ calculated static COHSEX self energy matrix elements (real and imaginary parts as separate points) for the training and test sets.
\textbf{b}, Test-set QP energies from the diagonal elements ($G_0W_0$) and from diagonalization of the QP Hamiltonian ($G_1W_0$). Inset: predicted vs.\ calculated $G_1W_0-G_0W_0$.
\textbf{c}, \textbf{d}, Band structures of the test materials FeZrCl$_6$ (\textbf{c}) and CuCl (\textbf{d}) from DFT (black), calculated $G_1W_0$ (blue, GW), and predicted $G_1W_0$ (orange dashed, GW-ML).}
\label{fig2}
\end{figure}

Fig.~\ref{fig2}a shows the predicted complex self-energy matrix elements against the calculated ones for the static COHSEX dataset. The model reaches $R^{2}=0.99998$ on the training set and $R^{2}=0.99987$ on the test set, demonstrating high-fidelity prediction of the self energy matrix across materials. The QP energies obtained from this matrix are also accurate (Fig.~\ref{fig2}b): the one-shot energies from its diagonal, $G_0W_0$, and the eigenvalues of the full QP Hamiltonian after diagonalizing, $G_1W_0$, both have a mean absolute error (MAE) of 29~meV.
The near equality of these two errors shows that diagonalization does not degrade the prediction.  The inset of Fig.~\ref{fig2}b shows the contribution of the off-diagonal self energy by plotting $G_1W_0-G_0W_0$ for each state, which the model reproduces with $R^{2}=0.95$ and an MAE of 4~meV, compared with a mean magnitude of 21 meV. Diagonalization also yields the QP wavefunctions, whose overlaps with the KS states are predicted with $R^{2}=0.978$ (see SI).
The predicted self energy also reproduces the QP dispersion across the Brillouin zone. Figures~\ref{fig2}c,d show the $G_1W_0$ band structures of the two test materials whose $GW$ corrections have the largest variation across band index and $\mathbf{k}$, respectively. In FeZrCl$_6$ the lowest valence band moves 1.2~eV more than the valence bands above it, and in CuCl the correction of a single band changes by up to 1~eV along $\mathbf{k}$.
The model reproduces both bandstructure with MAE of 38 and 12~meV, respectively. The $GW$ corrections are predicted on the $5\times5\times1$ input grid and interpolated to a dense path. More examples, including the material NaCNO with the largest MAE=174meV, are given in the SI. Once the DFT calculation is finished, the model predicts the self energy matrix of a material in about 3~s on one GPU node, when the first principle calculation takes on average 42 min on 256 CPUs

\begin{figure}
\centering
\includegraphics[width=\columnwidth]{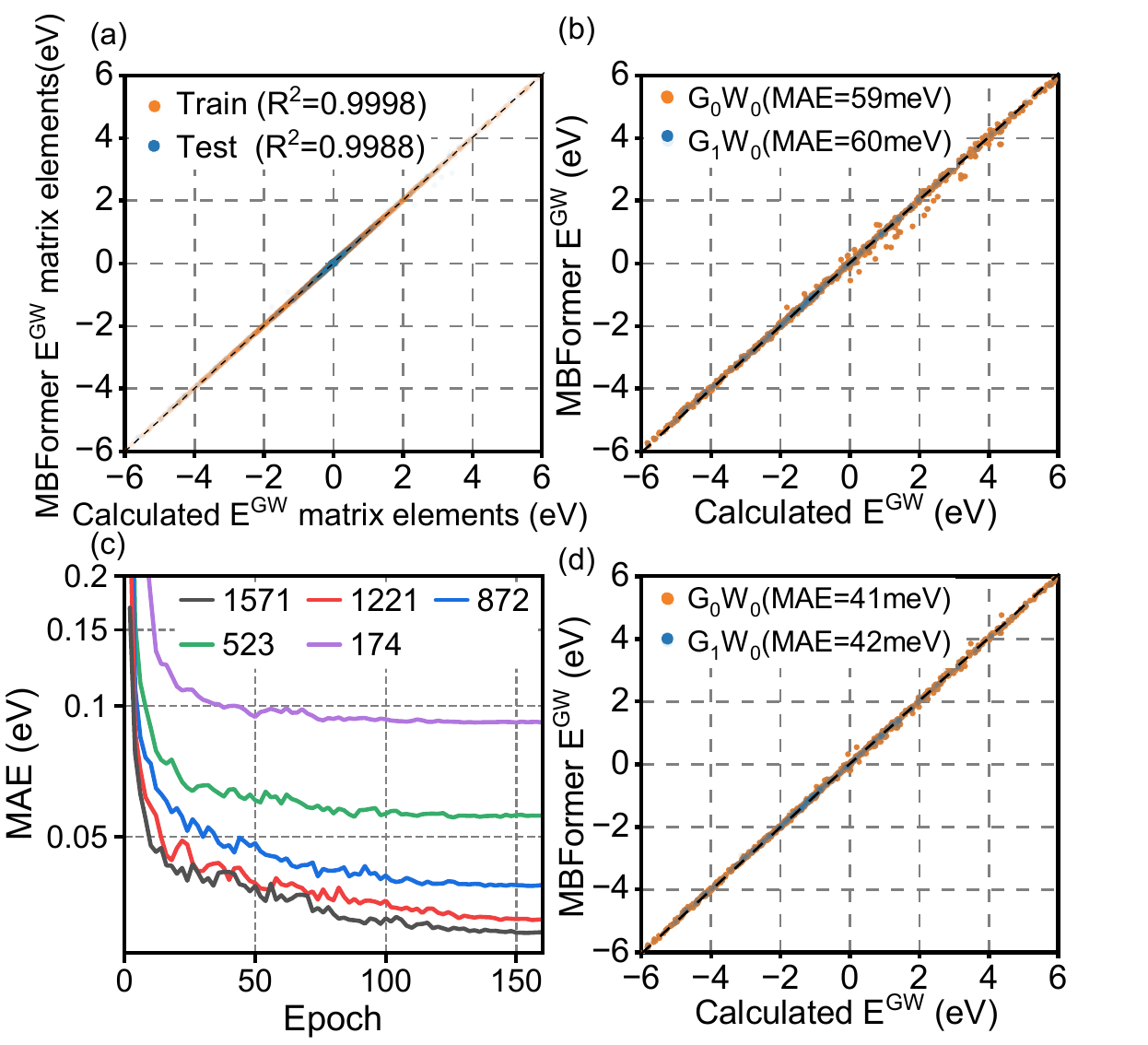}
\caption{\textbf{Prediction of GPP self energy matrices and data efficiency.}
\textbf{a}, Same as Fig.~\ref{fig2}a for the GPP self-energy matrix elements from the GPP dataset.
\textbf{b}, Same as Fig.~\ref{fig2}b for the GPP test set.
\textbf{c}, Test-set $G_1W_0$ MAE vs.\ training epoch for static COHSEX models trained on subsets of 174 to 1571 training materials and evaluated on the fixed set of 174 test materials.
\textbf{d}, Same as \textbf{b} for a model pretrained on the static COHSEX dataset and fine-tuned on the GPP training set for 20 epochs.}
\label{fig3}
\end{figure}

Figures~\ref{fig3}a and \ref{fig3}b show the corresponding results for the GPP dataset. Despite the smaller dataset, the model still attains $R^{2}=0.9998$ and $0.9988$ for the training and test sets, respectively, with corresponding MAEs of 59~meV for the one-shot $G_0W_0$ energies and 60~meV for the $G_1W_0$ energies. The larger errors compared with the static COHSEX prediction are consistent with the reduced amount of training data. As shown in Fig.~\ref{fig3}c for the static COHSEX dataset, the test-set $G_1W_0$ MAE decreases monotonically with the number of training materials, from 92~meV for 174 training materials to 30~meV for 1571. We therefore expect the GPP prediction accuracy to approach the static COHSEX level once more labels are generated.
If the model learns transferable structure in the mapping from mean-field states to the self energy, rather than correlations specific to the static COHSEX labels, pretraining on COHSEX should improve its prediction of the GPP self energy. 
Indeed, we find that fine-tuning the static COHSEX model on the GPP training set for only 20 epochs reduces the $G_0W_0/G_1W_0$ MAEs to 41/42~meV (Fig.~\ref{fig3}d), about 30\% lower than training from scratch. This setting reflects the practical situation in which inexpensive static COHSEX labels are available for all materials and the more expensive dynamically screened labels only for some of them.

\begin{figure}
\centering
\includegraphics[width=\columnwidth]{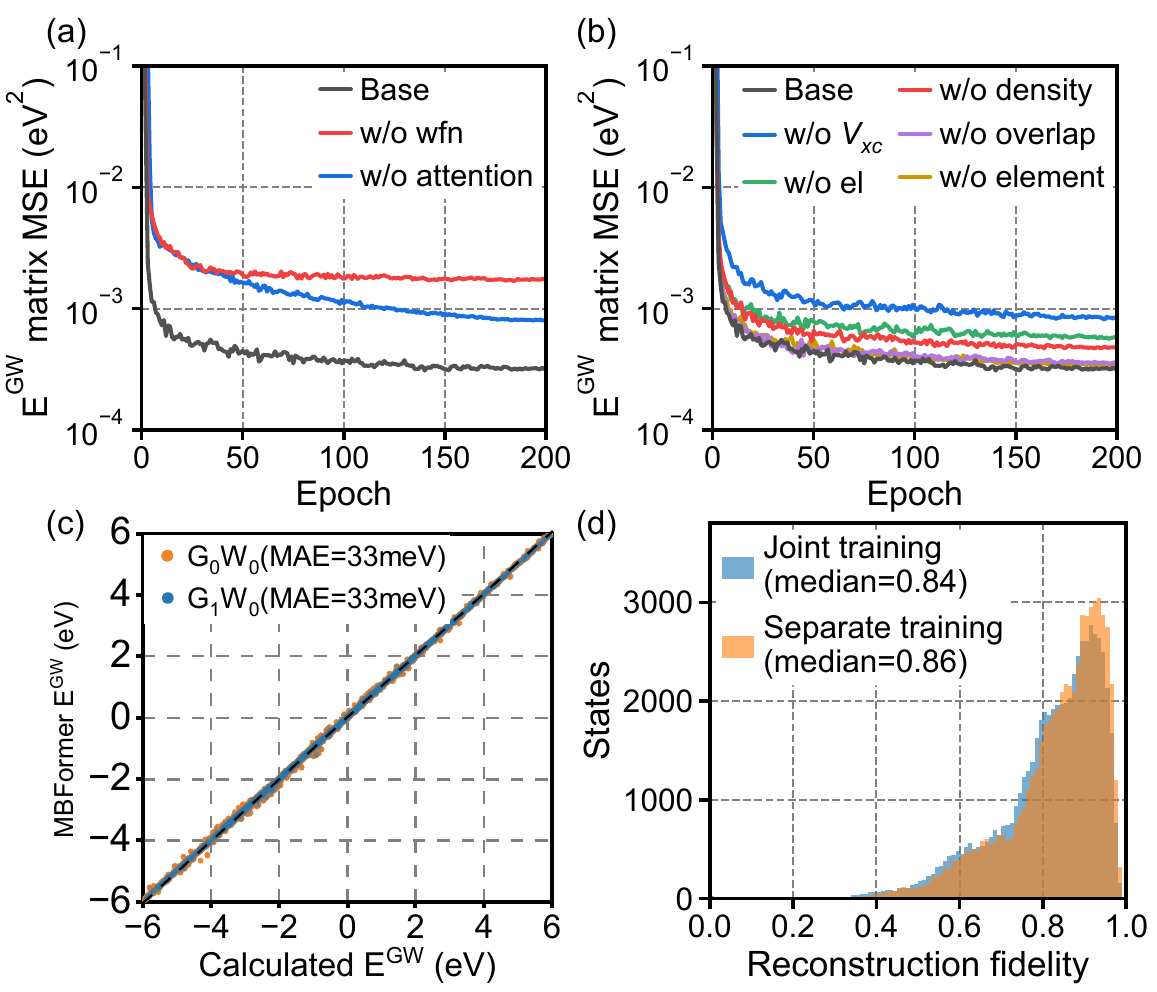}
\caption{\textbf{Ablation studies and joint versus separate training.}
\textbf{a}, Test set MSE of the static COHSEX self-energy matrix vs.\ epoch for the full model (Base), without the wavefunction input (w/o wfn), and with the transformer layers replaced by a MLP (w/o attention).
\textbf{b}, Same as \textbf{a} without the charge density, $V_{xc}$, the KS energies, the overlap descriptors, or the chemical element labels.
\textbf{c}, Same as Fig.~\ref{fig2}b with the autoencoder pretrained on reconstruction alone and frozen (separate training).
\textbf{d}, Reconstruction fidelity $\lvert\langle c\vert c^{\mathrm{rec}}\rangle\rvert/(\lvert c\rvert\,\lvert c^{\mathrm{rec}}\rvert)$ of the test states on the P grid for joint and separate training.}
\label{fig4}
\end{figure}

To identify the essential ingredients of the model, we perform ablation studies on the static COHSEX dataset, in which one input or component is removed and the model is retrained with otherwise identical settings (Figs.~\ref{fig4}a,b). We first test the two core components of the model: the wavefunction input and the transformer attention that models the many-body interaction. Removing all wavefunction inputs and the whole autoencoder structure increases the MSE of the test-set self-energy matrix by a factor of 5.5 (Fig.~\ref{fig4}a). Replacing the transformer layers by a multilayer perceptron, which removes all communication between band pairs and $\mathbf{k}$ points, increases the error by a factor of 2.5. Both the wavefunction information and the transformer attention between band pairs and $\mathbf{k}$ points are therefore indispensable. Among the other physical inputs (Fig.~\ref{fig4}b), removing $V_{xc}$, which then has to be learned as part of $\Sigma$, has the largest effect (factor 2.6), followed by the KS energies (1.8) and the density (1.5), while removing the overlap descriptors or the chemical element labels increases the error by 14\% and 9\%, respectively.

Next, using the static COHSEX dataset, we examine the effect of joint training on the accuracy of the $GW$ predictions. When the autoencoder is first trained on wavefunction reconstruction alone and then frozen before training the transformer, the model still predicts the self energy matrix elements with high fidelity $R^{2}=0.99983$. However, the QP-level MAEs increase to 33~meV for both $G_0W_0$ and $G_1W_0$ (Fig.~\ref{fig4}c), compared with 29~meV for joint training (Fig.~\ref{fig2}b). This difference does not arise from poor wavefunction reconstruction: the frozen autoencoder reconstructs the wavefunctions of test materials with a median fidelity of 0.86, slightly higher than the 0.84 of a jointly trained autoencoder (Fig.~\ref{fig4}d). Joint training instead lets the self energy loss reshape the latent space toward features relevant to the $GW$ prediction, at the cost of only a slight reduction in reconstruction fidelity.

\begin{figure}
\centering
\includegraphics[width=\columnwidth]{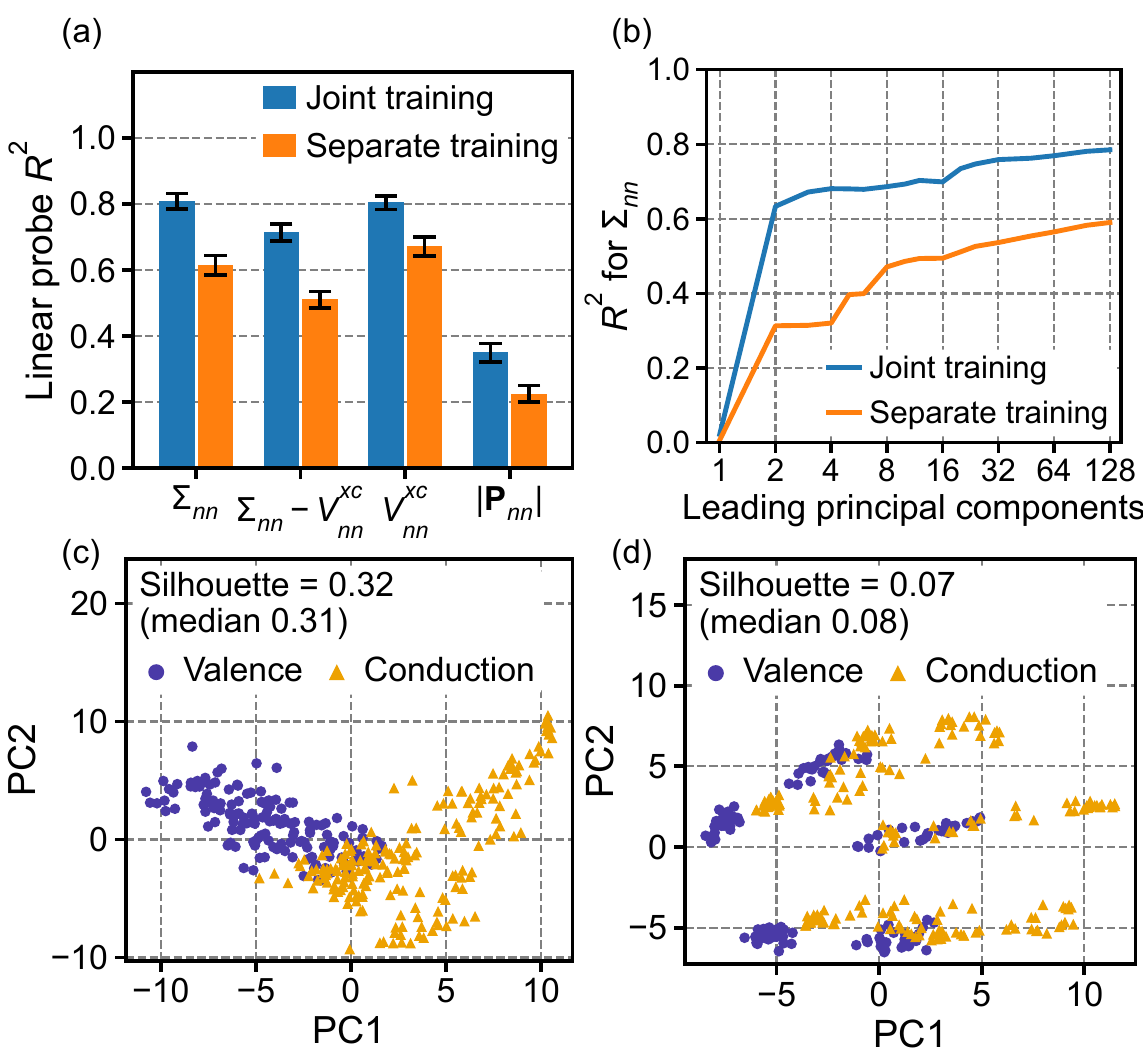}
\caption{\textbf{Joint training reshapes the latent into a more physically structured code.}
\textbf{a}, Test-set ridge-probe $R^{2}$ for $\Sigma_{nn}$, $\Sigma_{nn}-V^{xc}_{nn}$, $V^{xc}_{nn}$, and $\lvert\mathbf{P}_{nn}\rvert$; error bars are 95\% bootstrap intervals over test materials.
\textbf{b}, $R^{2}$ for $\Sigma_{nn}$ vs.\ the number of leading principal components retained; the full latents reach 0.81 (joint) and 0.61 (separate).
\textbf{c},\textbf{d}, Latents of the states of a representative test monolayer projected onto their two leading principal components and colored by occupation, for joint (\textbf{c}) and separate (\textbf{d}) training. The valence-conduction silhouette score is $0.32$ vs.\ $0.07$ (test median $0.31$ vs.\ $0.08$; joint larger in $88\%$ of test materials).}
\label{fig5}
\end{figure}

To identify what joint training changes in the representation itself, we probe the compressed latent directly (Fig.~\ref{fig5}). First, a linear ridge probe predicts the many-body targets more accurately from the jointly trained latent than from the separately trained one: $\Sigma_{nn}$ and the QP correction $\Sigma_{nn}-V^{xc}_{nn}$ have $R^{2}=0.81$ and 0.72, in the joint case compared with 0.61 and 0.51 in the separately trained case(Fig.~\ref{fig5}a). To test whether joint training also produces a more informative representation of the underlying KS states, we apply the same linear probe to properties of the KS states themselves. These are also predicted more accurately from the joint latent:  $V^{xc}_{nn}$ (0.80 vs.\ 0.67) and the magnitude of the momentum matrix element $\lvert\mathbf{P}_{nn}\rvert$ (0.35 vs.\ 0.22). Additionally, we find that the relevant information is more concentrated in the leading directions of the jointly trained representation. Figure~\ref{fig5}b shows the $R^2$ of $\Sigma_{nn}$ as a function of the number of leading principal components used by the probe. The two leading directions of the joint latent already carry as much information about $\Sigma_{nn}$ as the entire separate latent, and the joint curve stays above the separate one as more components are included. Finally, joint training produces a clearer separation between valence and conduction states in the two leading principal components of the latents of each material (Figs.~\ref{fig5}c,d). This is a physically relevant distinction because the self energy correction differs systematically between occupied and empty states. Together, these results show that the self energy objective reorganizes the wavefunction representation so that information relevant to the downstream $GW$ task is more accessible and more compactly encoded than in a latent trained for reconstruction alone.

In summary, we have demonstrated the $\mathbf{k}$-resolved prediction of the complex $GW$ self energy matrix across chemically and structurally diverse 2D semiconductors. Our model achieves a new state-of-the-art for the cross-material prediction of quasiparticle energiesin solids, and it is the first to do it with the full self energy matrix. The approach combines a phase-aware representation of the complex KS wavefunctions on a common physical-momentum grid with gauge augmentation and a factorized transformer that reflects the band and momentum structure of the $GW$ self energy. Joint end-to-end training further allows the many-body objective itself to shape the wavefunction representation, producing a latent space in which information relevant to both the KS states and the self energy is more directly accessible.
The predicted matrices retain their accuracy upon diagonalization and yield QP wavefunctions as well as energies, opening a route to accelerating calculations that require the full QP basis, including self-consistent $GW$, $GW$ perturbation theory for electron-phonon interactions~\cite{Li2019GWPT}, and nonequilibrium many-body dynamics. We show that the framework carries over from static COHSEX to dynamically screened GPP self energies without redesign and that pretraining on inexpensive static COHSEX labels improves the dynamical model with minimal additional training. More broadly, these results move machine learning of $GW$ beyond scalar QP corrections toward learning the many-body operator itself, giving access to both quasiparticle energies and wavefunctions in real materials.

\begin{acknowledgments}
This work was primarily supported by the U.S. Department of Energy, Office of Science, Basic Energy Sciences under Award No. DE-SC0026655.
Development of the BerkeleyGW code was supported by Center for Computational Study of Excited-State Phenomena in Energy Materials (C2SEPEM) at the Lawrence Berkeley National Laboratory, funded by the U.S. Department of Energy, Office of Science, Basic Energy Sciences, Materials Sciences and Engineering Division, under Contract No. DE-AC02-05CH11231. 
The calculations used resources of the National Energy Research Scientific Computing (NERSC), a Department of Energy, Office of Science User Facility, operated under contract number DE-AC02-05CH11231, under award numbers BES-ERCAP-0031507 and BES-ERCAP-0027380.
An award of computer time was provided by the U.S. Department of Energy’s (DOE) Innovative and Novel Computational Impact on Theory and Experiment (INCITE) Program. This research used supporting resources at the Argonne and the Oak Ridge Leadership Computing Facilities. The Argonne Leadership Computing Facility at Argonne National Laboratory is supported by the Office of Science of the U.S. DOE under Contract No. DE-AC02-06CH11357 . The Oak Ridge Leadership Computing Facility at the Oak Ridge National Laboratory is supported by the Office of Science of the U.S. DOE under Contract No. DE-AC05-00OR22725.
This material is based upon research conducted by D.Y.Q. as an Alfred P. Sloan Research Fellow under Grant No. FG-2026-79515.

\end{acknowledgments}

\emph{Data and code availability}.
The numerical data underlying the figures and the trained model checkpoints and inference code used to generate the results reported in this work will be deposited in the Materials Data Facility. Other data are available from the authors upon reasonable request.

\bibliography{references}

\end{document}


\title{
Supplemental Information for \textit{Machine Learning the \texorpdfstring{$GW$}{GW} Self Energy Operator for Real Materials}
}
\author{Xian Xu}
\affiliation{Department of Applied Physics, Yale University, New Haven, CT, 06511, USA}
\author{Bowen Hou}
\affiliation{Department of Materials Science, Yale University, New Haven, CT, 06511, USA}
\author{Chengyan Zhang}
\affiliation{Department of Applied Physics, Yale University, New Haven, CT, 06511, USA}
\author{Diana Y. Qiu}
\email{diana.qiu@yale.edu}
\affiliation{Department of Materials Science, Yale University, New Haven, CT, 06511, USA}

\date{\today}

\maketitle

\begin{figure}[p]
    \centering
    \includegraphics[width=\textwidth]{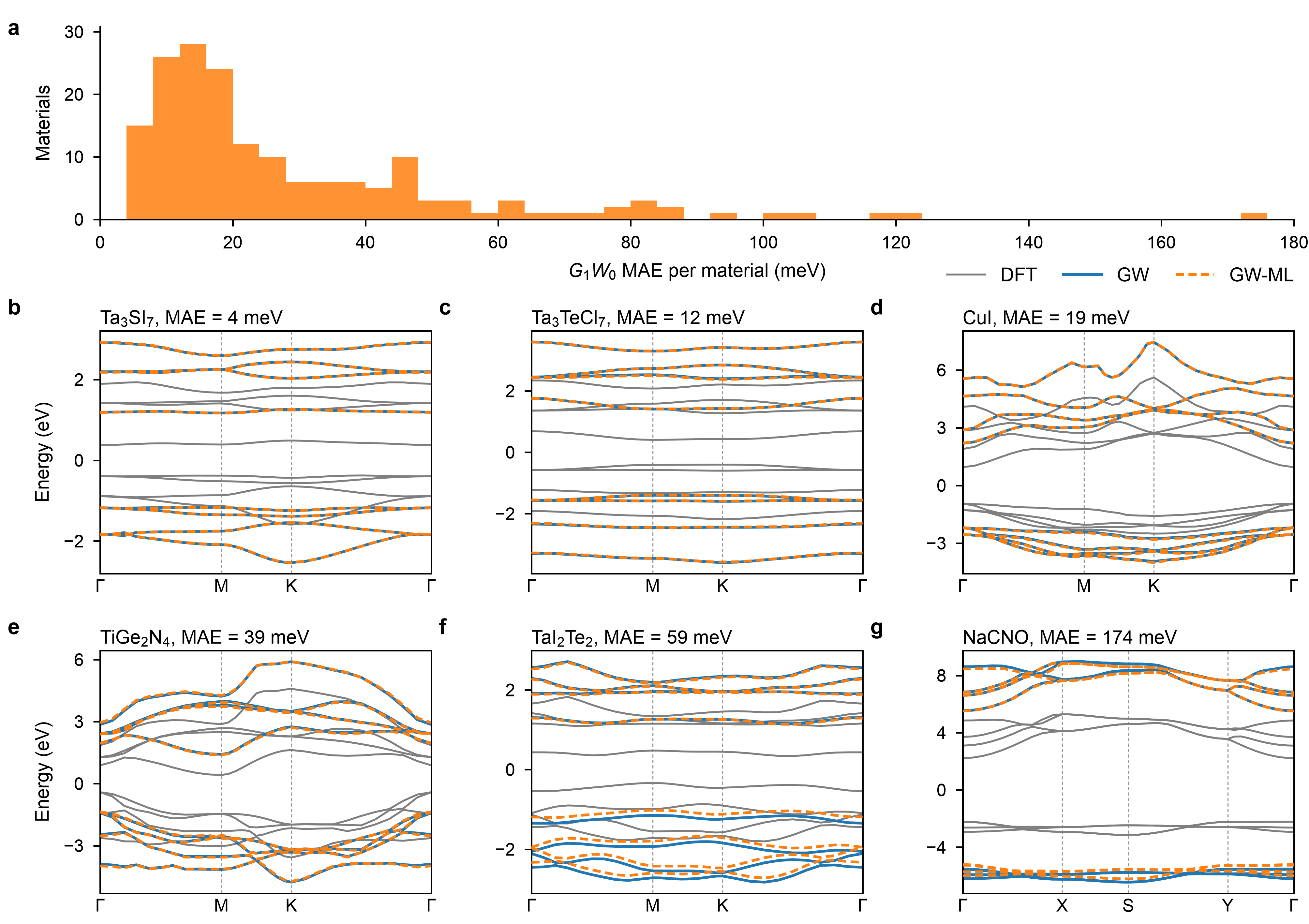}
    \caption{\textbf{Errors and band structures of static COHSEX test materials.}
    \textbf{a}, Distribution of the $G_1W_0$ MAE for each of the 174 test materials for the model shown in Fig.~2 of the main text. The median MAE is 19~meV, and the mean MAE is 29~meV, with NaCNO having the largest MAE (174~meV). The test-set MAE for the smallest band gap on the $5\times5\times1$ grid is 59~meV for this model, 120~meV for the GPP model shown in Fig.~3b of the main text, and 81~meV for the finetuned model of Fig.~3d of the main text. \textbf{b} to \textbf{g}, Band structures for a range of test materials with different MAEs at the DFT level (black) calculated $G_1W_0$ level (blue), and the predicted $G_1W_0$ (GW-ML) level (orange dashed). \textbf{g} is the bandstructure of the material with the largest MAE in our dataset. As in Figs.~2c,d of the main text, the energies of the four highest valence and four lowest conduction bands are interpolated from the $5\times5\times1$ grid to the fixed path using the dual grid interpolation scheme implemented in BerkeleyGW~\cite{Deslippe2012BerkeleyGW}}.
    \label{fig:bands}
\end{figure}
\clearpage

\begin{figure}[p]
    \centering
    \includegraphics[width=0.5\textwidth]{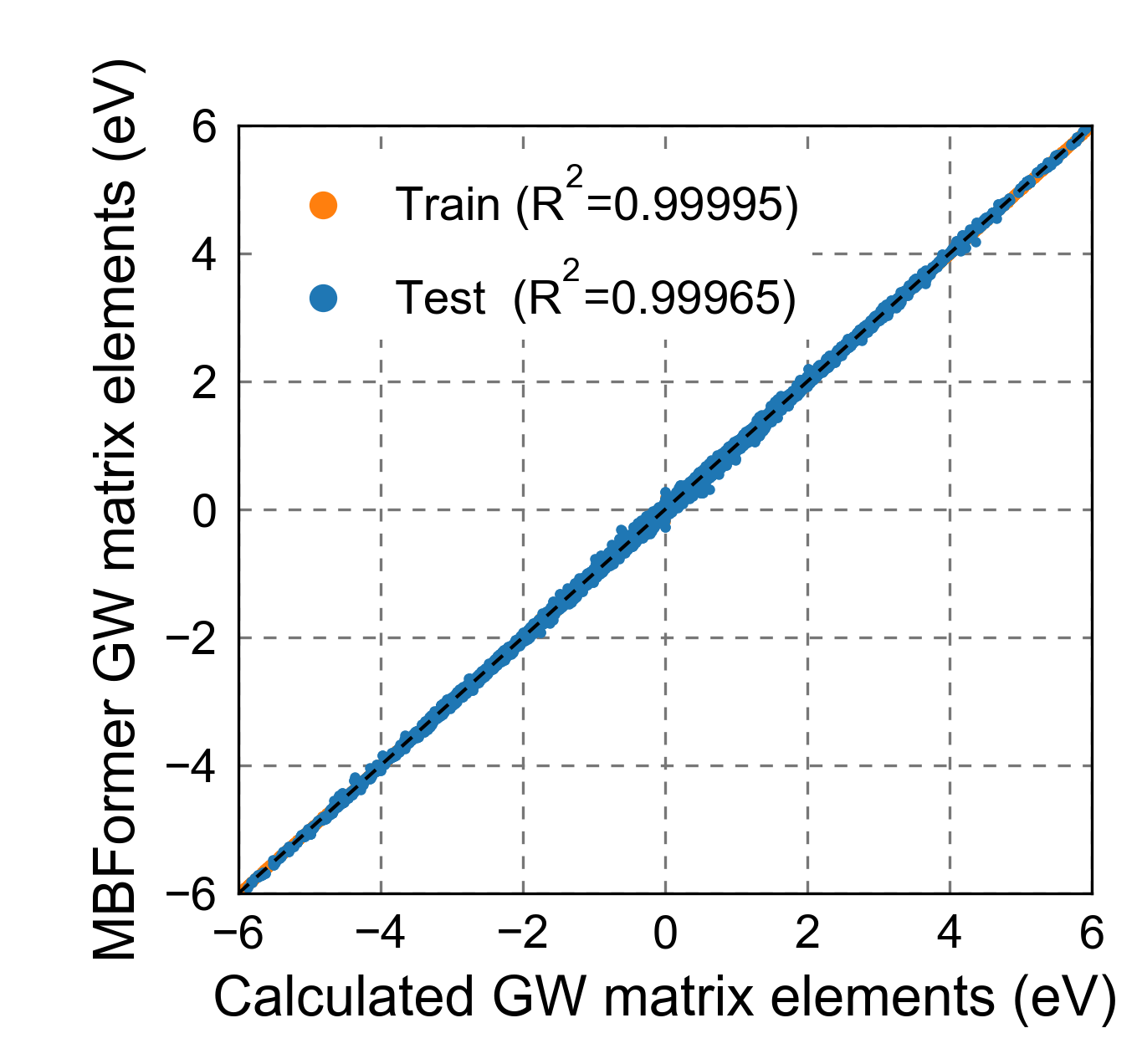}
    \caption{\textbf{Off-diagonal self energy matrix elements.} Same as Fig.~2a of the main text for the off-diagonal elements $\Sigma_{mn\mathbf{k}}$ with $m\neq n$ only. The test set MAE is 5.0~meV per real or imaginary part.}
    \label{fig:offdiag}
\end{figure}
\clearpage

\section{Note 1: Datasets and computational parameters}

The static COHSEX dataset contains 1745 2D semiconductors with structures obtained from C2DB~\cite{Haastrup2018C2DB,Gjerding2021C2DB}, with 2 to 14 atoms per unit cell and PBE~\cite{Perdew1996PBE} band gaps of 0.20 to 8.05~eV. The GPP dataset contains 433 hexagonal 2D semiconductors with 2 to 6 atoms per unit cell and PBE band gaps of 0.22 to 7.44~eV. For each dataset, 90\% (10\%) of the materials are randomly selected for training (test), resulting in 1571 (174) static COHSEX and 389 (44) GPP materials.

All DFT calculations are performed using Quantum ESPRESSO~\cite{Giannozzi2009} with the PBE functional and scalar relativistic optimized norm-conserving Vanderbilt pseudopotentials~\cite{Hamann2013ONCV}, without spin polarization or spin-orbit coupling. The cutoffs of the wavefunctions and the charge density are 50 and 200~Ry, the cell height is 12~\AA{}, and the $\mathbf{k}$ grid is $5\times5\times1$ without symmetry reduction. All $GW$ calculations are performed using BerkeleyGW~\cite{Deslippe2012BerkeleyGW} with slab Coulomb truncation~\cite{IsmailBeigi2006}. The energy cutoff for the planewave components of the screened Coulomb interaction is 20~Ry for the static COHSEX dataset and (25~Ry for the generalized plasmon-pole model (GPP)~\cite{Hybertsen1986}). Stochastic pseudobands~\cite{Altman2024Pseudobands} are used to accelerate the sums over empty bands, and the nonuniform neck subsampling scheme~\cite{Jornada2017NNS} is used to converge the screening at small $\mathbf{q}$. On 256 MPI tasks with two OpenMP threads each, the $GW$ calculation takes on average 42~min for the static COHSEX dataset and 87~min for the GPP dataset. For the 411 materials contained in both datasets, it takes 10~min for static COHSEX and 88~min for GPP(\ref{eq:gpp}).

The target is the self energy matrix in the basis of the six highest valence and eight lowest conduction bands (resulting in a matrix of size $14\times14$) for the static COHSEX dataset and of five valence and four conduction bands (resulting in a matrix of size $9\times9$) for the GPP dataset. The static COHSEX self energy is Hermitian by construction. The GPP self energy depends on the energy at which it is evaluated, so its off-diagonal elements are evaluated at the Kohn-Sham (KS) energies of both states and symmetrized,
\begin{equation}
\Sigma_{mn\mathbf{k}}=\tfrac{1}{2}\bigl[\Sigma_{mn\mathbf{k}}(\varepsilon_{m\mathbf{k}})+\Sigma_{mn\mathbf{k}}(\varepsilon_{n\mathbf{k}})\bigr],
\label{eq:gpp}
\end{equation}
analogous to the procedure in quasiparticle self-consistent $GW$~\cite{Faleev2004,vanSchilfgaarde2006QSGW}.
\clearpage

\section{Note 2: The P grid}

\begin{figure}[p]
    \centering
    \includegraphics[width=0.93\textwidth]{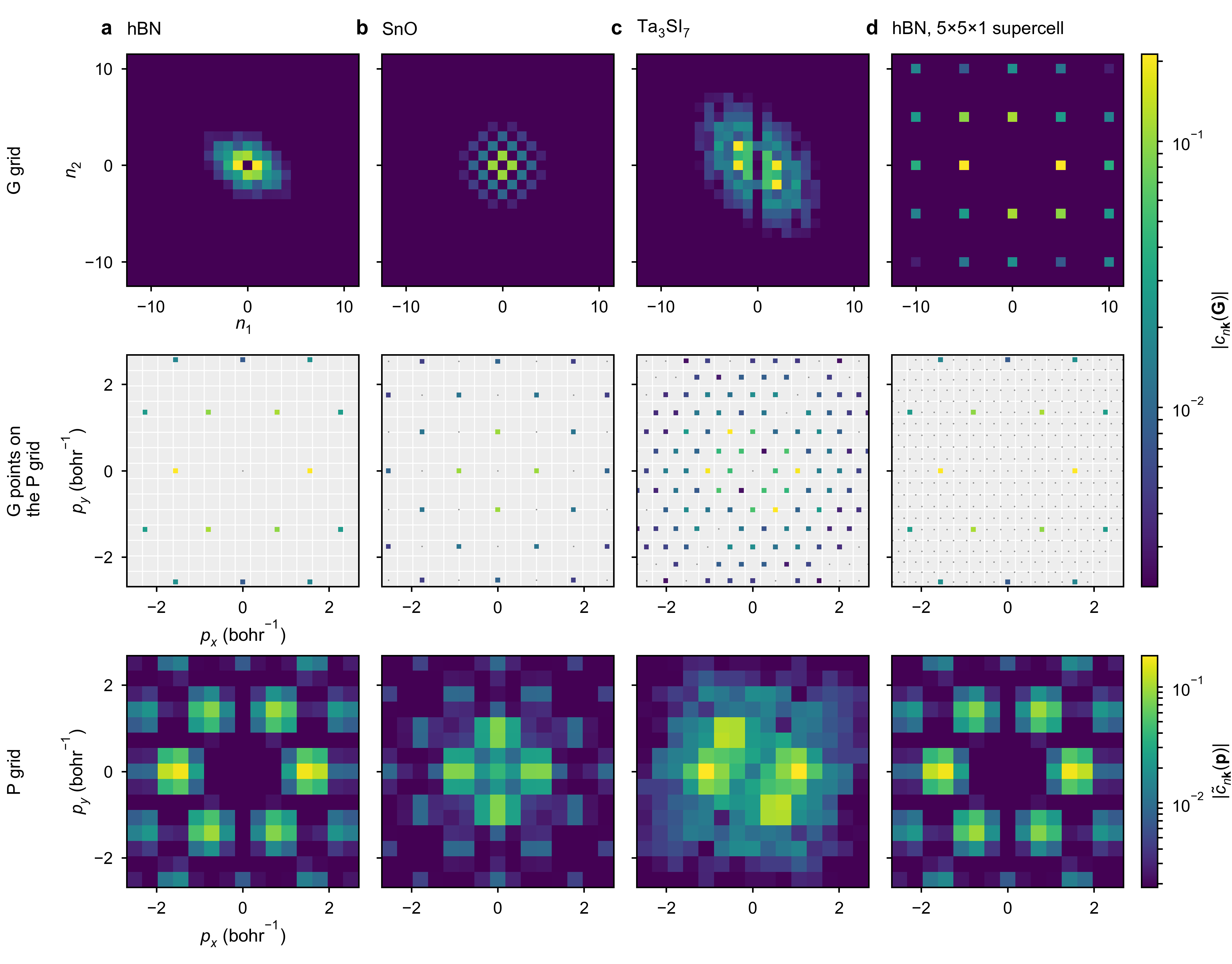}
    \caption{\textbf{The same states on the G grid and on the P grid.}
    Coefficients $\lvert c_{n\mathbf{k}}\rvert$ of one KS state at $\Gamma$: \textbf{a}, the highest valence states of hexagonal boron nitride (hBN), which has a hexagonal lattice; \textbf{b}, the highest valence state of SnO, which has a square lattice; \textbf{c}, the sixth highest valence state of Ta$_3$SI$_7$, which has a hexagonal lattice with a much larger lattice constant than hBN; \textbf{d}, the same state as in \textbf{a} from a $5\times5\times1$ supercell of hBN. Top, the G grid at $G_z=0$. Middle, the same coefficients at their momenta $\mathbf{p}=\mathbf{k}+\mathbf{G}$, on the P grid (white lines). Bottom, the P grid after the Gaussian broadening}.
    \label{fig:pgrid}
\end{figure}

The plane-wave coefficients $c_{n\mathbf{k}}(\mathbf{G})$ are naturally stored on a reciprocal-lattice grid indexed by $\mathbf{G}$. Across different crystals, however, the same grid index corresponds to different physical momenta because the reciprocal-lattice vectors depend on the lattice geometry and lattice constants. To give the coefficient array a common physical meaning across materials, we instead represent each coefficient by its Cartesian momentum $\mathbf{p}=\mathbf{k}+\mathbf{G}$, which we refer to as the \textbf{P} grid. Each coefficient is placed at this physical momentum and distributed over nearby nodes of a fixed Cartesian momentum grid using a Gaussian kernel with a broadening of two thirds of the node spacing (0.23~bohr$^{-1}$ in plane and 0.12~bohr$^{-1}$ out of plane). We use $(n_x,n_y,n_z)=(16,16,36)$ nodes.

Because this resampling interpolates the original coefficients onto a fixed grid, we test how much information is lost in the transformation. We reconstruct the original coefficients from the P-grid representation by solving the inverse resampling problem using least squares with Tikhonov regularization. The reconstructed states have a median fidelity of 0.990, showing that the fixed-grid representation retains nearly all of the information in the original plane-wave coefficients.

The resulting representation is illustrated in Fig.~\ref{fig:pgrid}. On the original $\mathbf{G}$ grid, states from materials with different lattice geometries or lattice constants occupy different index patterns even when their physical momentum content is similar. On the P grid, each node corresponds to the same physical momentum in every material. In particular, the primitive-cell and $5\times5\times 1$ supercell representations of the same state in hexagonal boron nitride (hBN) map onto the same field on the P grid, demonstrating that the representation depends on physical momentum rather than the choice of reciprocal-lattice indexing.
\clearpage

\FloatBarrier
\section{Note 3: Input features}

\begin{figure}[tbp]
    \centering
    \includegraphics[width=\textwidth]{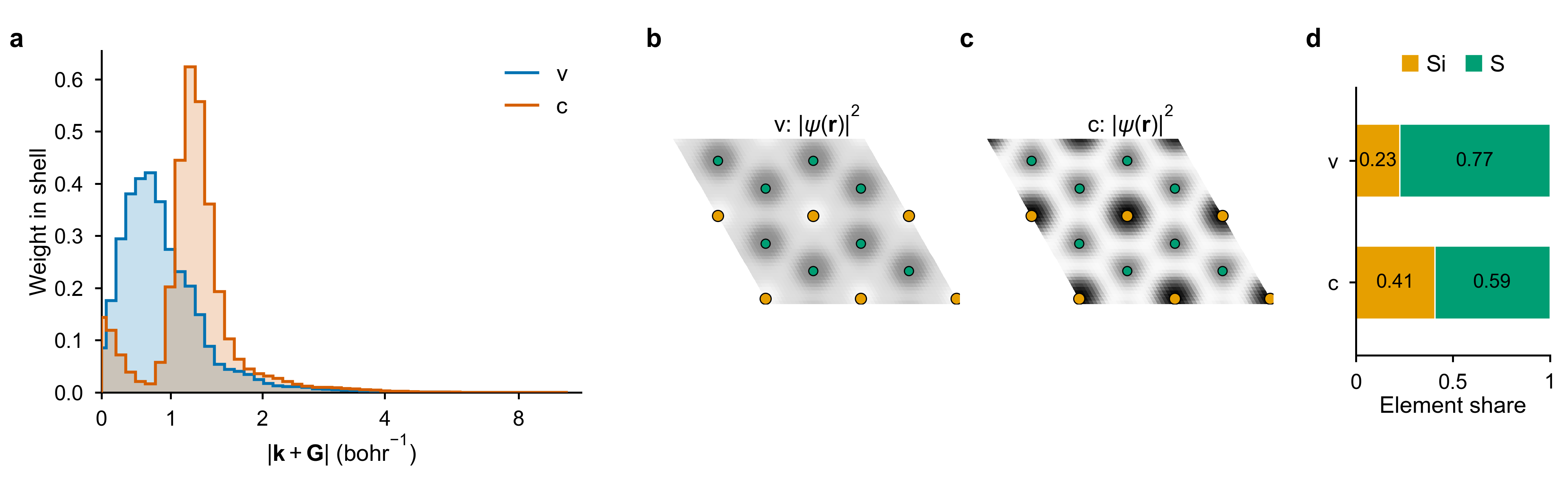}
    \caption{\textbf{State inputs.} Two states of the test material SiS$_2$ at $\Gamma$: the lowest valence state v and the lowest conduction state c. \textbf{a}, Radial encoding: the weight $\lvert c_{n\mathbf{k}}(\mathbf{G})\rvert^2$ collected in each of the 48 shells of $\lvert\mathbf{k}+\mathbf{G}\rvert$, one bin per shell. v has its weight at smaller momenta than c. \textbf{b}, \textbf{c}, $\lvert\psi_{n\mathbf{k}}(\mathbf{r})\rvert^2$ summed along $z$, with the Si (orange) and S (green) atoms. \textbf{d}, Element shares of v and c: c lives more on Si than v.}
    \label{fig:state_inputs}
\end{figure}

\begin{figure}[tbp]
    \centering
    \includegraphics[width=\textwidth]{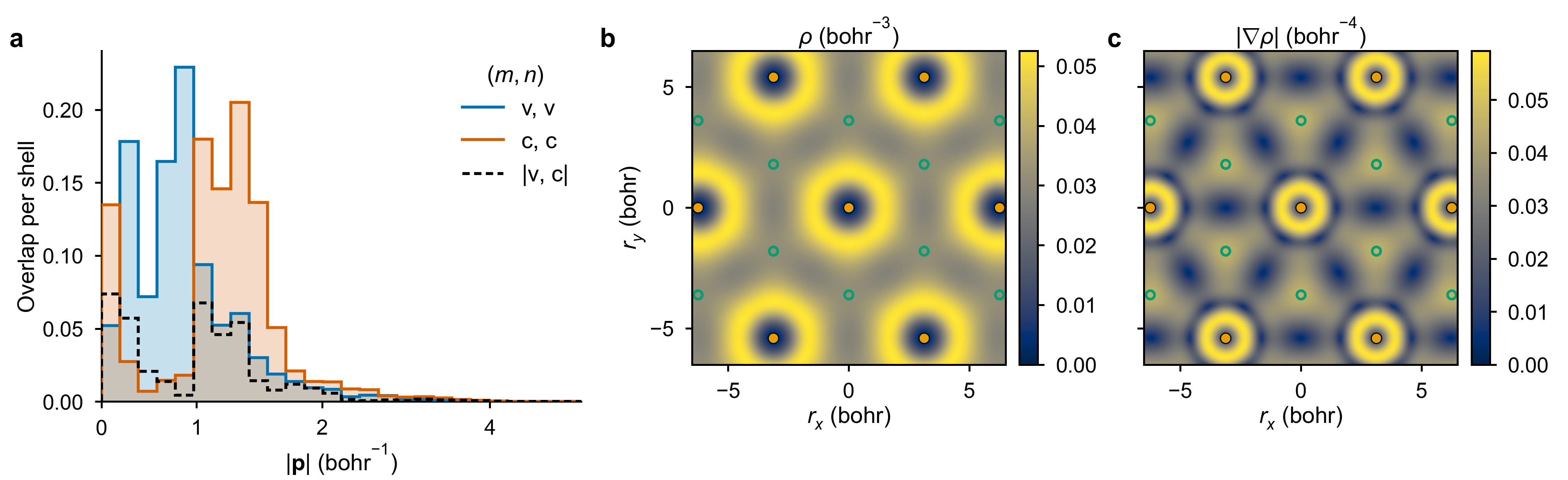}
    \caption{\textbf{Pair inputs} for the states v and c of Fig.~\ref{fig:state_inputs}. \textbf{a}, Overlaps $\mathcal{S}^{\nu}_{mn}$ resolved in momentum for $(m,n)=$ (v, v) and (c, c), and the magnitude for (v, c), one bin per shell $\nu$ of $\lvert\mathbf{p}\rvert$. For $m=n$, they give the momentum distribution of the state on the P grid and add up to one; the zigzag follows the number of nodes of the P grid in each shell. \textbf{b}, \textbf{c}, The ground state valence density $\rho$ and its gradient $\lvert\nabla\rho\rvert$ in the plane of the Si atoms (orange); the S atoms (green, open circles) lie above and below this plane.}
    \label{fig:pair_inputs}
\end{figure}

Besides the latent, each state carries its KS energy, its band indices, a radial encoding, and an element decomposition.

The radial encoding describes how the weight of a state is distributed over the momentum $\lvert\mathbf{k}+\mathbf{G}\rvert$. The weights $\lvert c_{n\mathbf{k}}(\mathbf{G})\rvert^2$ on the G grid are collected in 48 smooth Gaussian shells, evenly spaced in $\ln(1+\lvert\mathbf{k}+\mathbf{G}\rvert^2)$ (Fig.~\ref{fig:state_inputs}a), and a perceptron with two layers maps the 48 shell weights to 24 features. The positions and widths of the shells are trainable but stay close to these initial values.

The element decomposition describes on which atoms a state lives. Each point of the grid in real space is shared among the nearby atoms with Gaussian weights of width 1.2~bohr, and $\lvert\psi_{n\mathbf{k}}(\mathbf{r})\rvert^2$ is summed with these weights for each atom and then over the atoms of each element (Fig.~\ref{fig:state_inputs}b to d). All the elements' shares are mapped linearly to 16 features.

Each band pair carries the exchange-correlation matrix element $V_{xc,mn\mathbf{k}}$, momentum-resolved overlap descriptors, and charge-density overlaps, which all transform like $\Sigma_{mn\mathbf{k}}$ under Eq.~(1) of the main text (Fig.~\ref{fig:pair_inputs}). The overlap in momentum shell $\nu$ is defined as $S^\nu_{mn\mathbf{k}}$, the sum of $\tilde c^{*}_{m\mathbf{k}}\tilde c_{n\mathbf{k}}$ over the nodes of the P grid in this shell. The 26 shells are evenly spaced in $\ln(1+\lvert\mathbf{p}\rvert^2)$ (Fig.~\ref{fig:pair_inputs}a). Similarly, the charge density overlap $D_{mn\mathbf{k}}=\mel{m\mathbf{k}}{\rho}{n\mathbf{k}}$ is resolved over the regions of real space with given values of the ground state valence density $\rho$ (10 bins) and of its reduced gradient $s=\frac{\lvert\nabla\rho\rvert}{2(3\pi^2)^{1/3}\rho^{4/3}}$ (6 bins) (Fig.~\ref{fig:pair_inputs}b,c). The token $x_{mn\mathbf{k}}$ is a linear map of the normalized $h_{m\mathbf{k}}$, $h_{n\mathbf{k}}$, these pair-dependent quantities, and five energy features: $\varepsilon_{m\mathbf{k}}-\varepsilon_{n\mathbf{k}}$ and, for both states, the energy relative to the valence band maximum and to the conduction band minimum of the window.
\clearpage

\FloatBarrier
\section{Note 4: Quasiparticle wavefunctions}

\begin{figure}[tbp]
    \centering
    \includegraphics[width=0.5\textwidth]{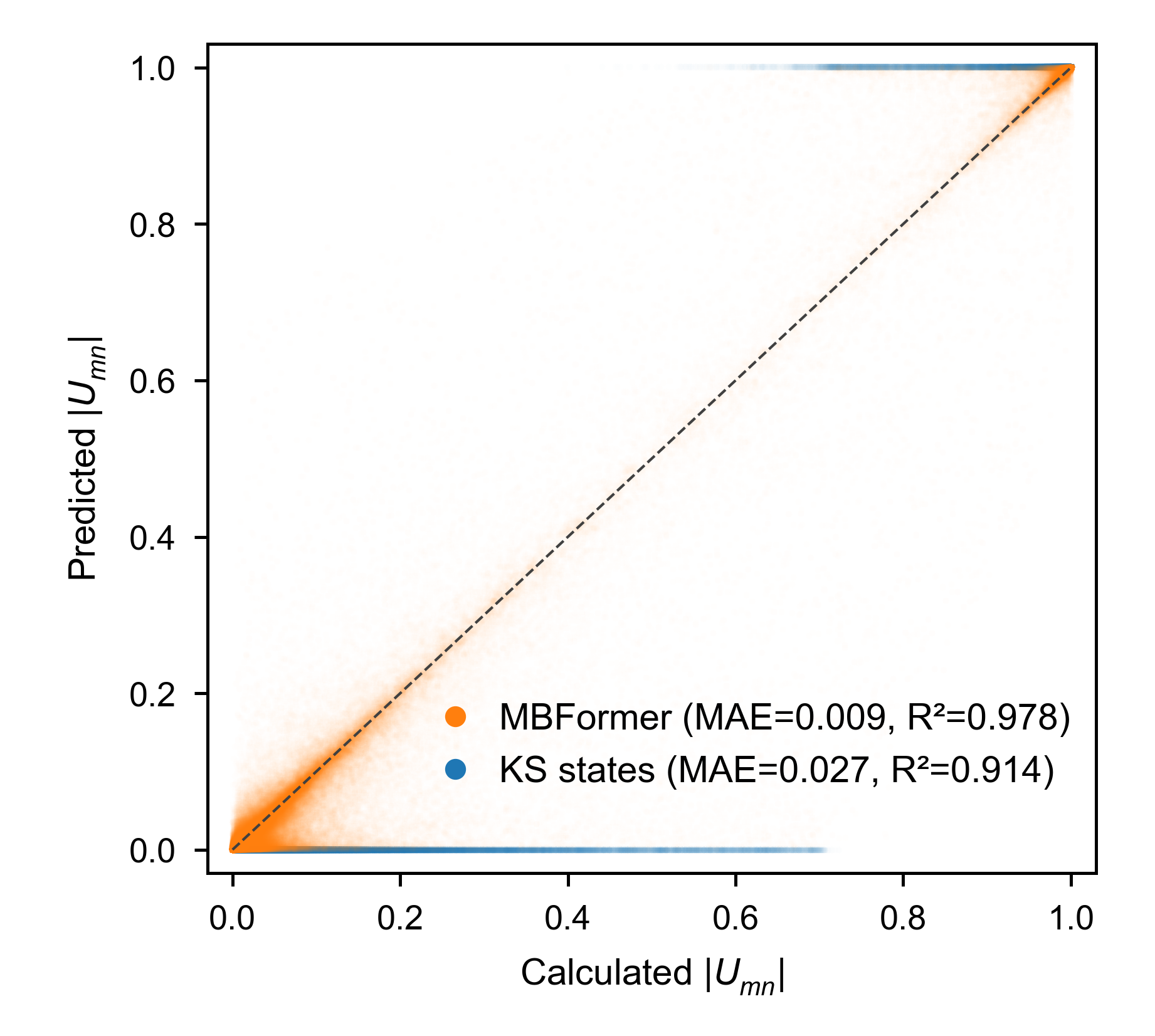}
    \caption{\textbf{Accuracy of the quasiparticle (QP) states.} Absolute values of the overlaps $U_{mn}$ between the KS states $m$ and the QP states $n$ of $H^{\mathrm{QP}}_{\mathbf{k}}$, predicted vs.\ calculated, for the 174 static COHSEX test materials, from MBFormer (orange) and with the KS states taken as the QP states (blue).}
    \label{fig:qp_wfn}
\end{figure}

Diagonalizing $H^{\mathrm{QP}}_{\mathbf{k}}$ writes each QP state $\ket{G_1,n}$ as a combination of the KS states $\ket{G_0,m}$, with overlaps $U_{mn}=\braket{G_1,n}{G_0,m}$. In the static COHSEX test set the QP states differ markedly from the KS states. The predicted $\lvert U_{mn}\rvert$ agree with the calculated ones with $R^{2}=0.978$ and an MAE of 0.009 (Fig.~\ref{fig:qp_wfn}), compared with $R^{2}=0.914$ and 0.027 when the KS states are taken as the QP states.
\clearpage

\section{Note 5: Autoencoder architecture}
\begin{figure}[tbp]
    \centering
    \includegraphics[width=0.25\textwidth]{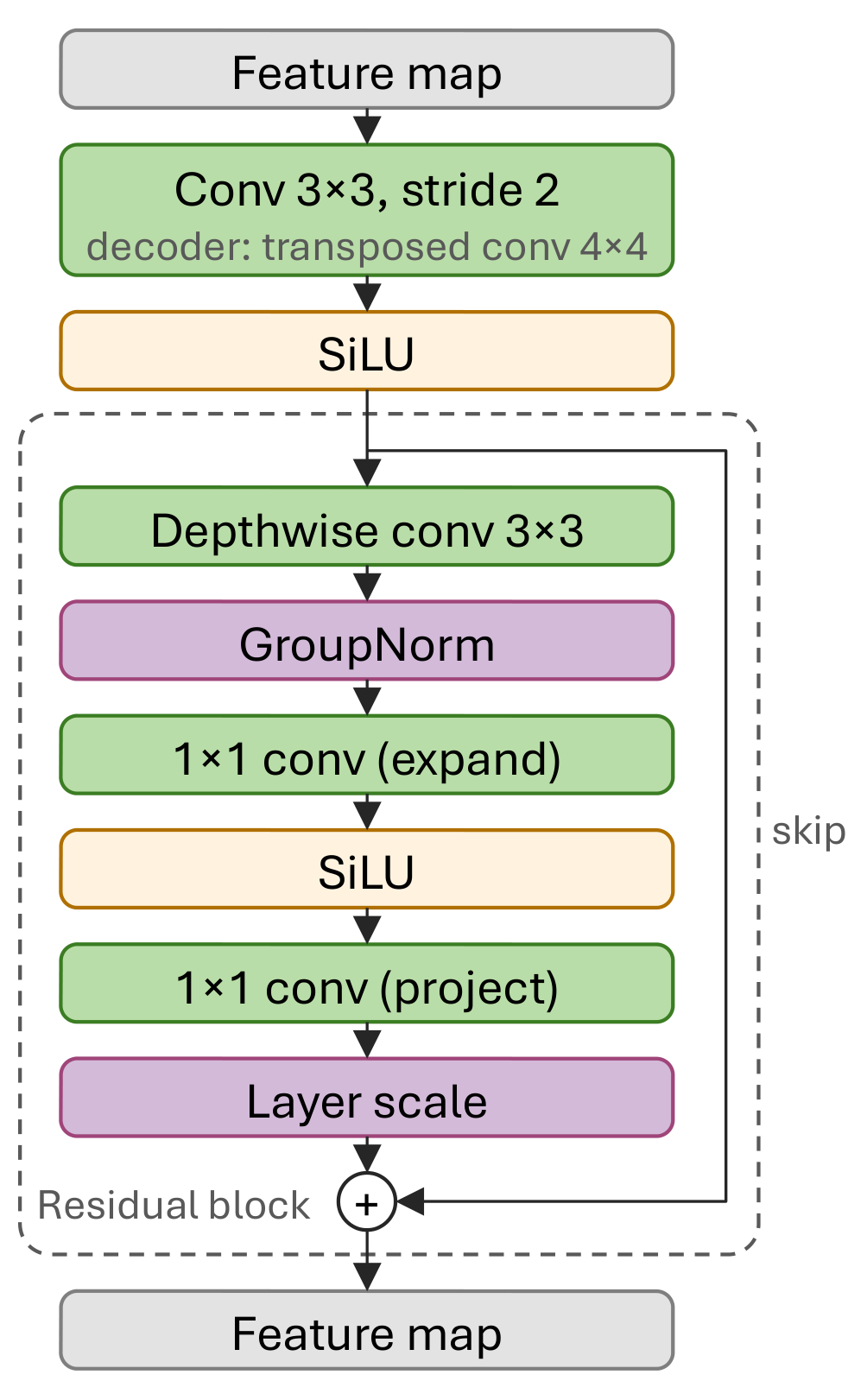}
    \caption{\textbf{Convolution block of the autoencoder.} Each encoder and decoder has three such blocks, from the $16\times16$ inplane nodes of the P grid to $2\times2$, and a linear map to connect the latent. The $p_z$ nodes enter as channels. }
    \label{fig:ae_stage}
\end{figure}

Our autoencoder is deterministic and contains no variational step. We treat the $16\times16$ inplane nodes of the P grid as an image and the 36 out-of-plane nodes as channels. Each encoder consists of three convolution blocks (Fig.~\ref{fig:ae_stage}) with 96, 192, and 384 channels, which progressively reduce the inplane grid to $2\times2$, followed by a linear layer that maps the result to the latent space. The amplitude and phase encoders produce the latents $z_a$ and $z_\phi$, each with 140 dimensions. The decoders mirror this structure with transposed convolutions. To keep the reconstructed amplitude nonnegative, the amplitude decoder ends with a softplus activation, while the phase decoder outputs two numbers per node that are normalized onto the unit circle to give $(\cos\hat\phi,\sin\hat\phi)$. Both branches are trained with the mean squared error of the reconstructed amplitudes and of the real and imaginary parts of the coefficients, multiplied by a weight $\lambda=5\times10^{-4}$.

\clearpage

\section{Note 6: Transformer architecture and training details}

We represent each band pair $(m,n)$ with $m\le n$ at each $\mathbf{k}$ point as one token, since the remaining elements follow from Hermiticity. This gives 105 tokens per $\mathbf{k}$ point for static COHSEX and 45 for GPP. Each token is built by concatenating the state vectors $h_{m\mathbf{k}}$ and $h_{n\mathbf{k}}$ with the pair features described in Note~3, and is then projected linearly to a dimension of 640. The transformer contains three factorized layers. Within each layer, attention is first applied along the $\mathbf{k}$ axis and then along the band-pair axis, after which a triangle update refines the tokens. All attention blocks use 8 heads with root mean square normalization~\cite{Zhang2019RMSNorm} and a SwiGLU feedforward network~\cite{Shazeer2020SwiGLU} with a hidden dimension of 2048. The triangle update couples the band pairs at the same $\mathbf{k}$ point through a sum over an intermediate band $l$,
\begin{equation}
x_{mn}\to x_{mn}+\tfrac{1}{2}W_o\,\mathrm{LN}(t_{mn}),\qquad t_{mn}=\sum_{l}\alpha_{ml}\beta_{ln},
\label{eq:triangle}
\end{equation}
where $\alpha$ and $\beta$ are gated linear projections of the tokens with 64 channels (the product is taken channel by channel), $\mathrm{LN}$ denotes layer normalization, and $W_o$ is a linear layer. This form is motivated by the band sum in $\mathbf{M}^{\dagger}W\mathbf{M}$ of Eq.~(2) of the main text. Finally, a small output head maps each token to the real and imaginary parts of $\Delta\Sigma_{mn\mathbf{k}}$. For $\mathcal{L}_{\Sigma}$ we use the mean absolute error of the complex matrix elements of $\Delta\Sigma$, and the total loss $\mathcal{L}=\mathcal{L}_{\Sigma}+\mathcal{L}_{\mathrm{rec}}$ is averaged over the materials in each training step. During training, we multiply every state by a random phase at each step and corotate $\Sigma$, $V_{xc}$, and both sets of overlaps according to Eq.~(1) of the main text. We use this gauge augmentation because imposing exact equivariance throughout the complex MBFormer architecture would require restricting every transformer operation. Our previous work showed that phase augmentation can impose an approximate $U(1)$ equivariance even without an explicitly equivariant architecture~\cite{Zhang2026}. We optimize all weight matrices of the linear layers with Muon~\cite{Jordan2024Muon} and the remaining parameters with Adam~\cite{Kingma2014Adam}. No weight decay is used. The learning rate follows a cosine decay from $10^{-4}$ to $10^{-6}$ over 200 epochs. Dropout with a rate of 0.1 is applied to the attention weights, the attention outputs, and the hidden layer of the SwiGLU network. All reported results use the model after the final training epoch.

\bibliography{references}